\documentclass[aps,prb,twocolumn,letterpaper,superscriptaddress,showpacs]{revtex4-1}
\usepackage{keyval}%
\usepackage{graphicx}
\usepackage{dcolumn}
\usepackage{bm}
\usepackage{color}
\usepackage{amsmath, amssymb}
\usepackage{soul}

\begin{document}

\title{Single-particle excitation of core states in epitaxial silicene}
\author{Chi-Cheng Lee}
\affiliation{Institute for Solid State Physics, The University of Tokyo, 5-1-5 Kashiwanoha, Kashiwa, Chiba 277-8581, Japan}%
\author{Jun Yoshinobu}
\affiliation{Institute for Solid State Physics, The University of Tokyo, 5-1-5 Kashiwanoha, Kashiwa, Chiba 277-8581, Japan}%
\author{Kozo Mukai}
\affiliation{Institute for Solid State Physics, The University of Tokyo, 5-1-5 Kashiwanoha, Kashiwa, Chiba 277-8581, Japan}%
\author{Shinya Yoshimoto}
\affiliation{Institute for Solid State Physics, The University of Tokyo, 5-1-5 Kashiwanoha, Kashiwa, Chiba 277-8581, Japan}%
\author{Hiroaki Ueda}
\affiliation{Institute for Solid State Physics, The University of Tokyo, 5-1-5 Kashiwanoha, Kashiwa, Chiba 277-8581, Japan}%
\author{Rainer Friedlein}
\affiliation{School of Materials Science, Japan Advanced Institute of Science and Technology (JAIST), 1-1 Asahidai, Nomi, Ishikawa 923-1292, Japan}
\affiliation{\textnormal{Current address}: Meyer Burger (Germany) AG, An der Baumschule 6-8, 09337 Hohenstein-Ernstthal, Germany}
\author{Antoine Fleurence}
\affiliation{School of Materials Science, Japan Advanced Institute of Science and Technology (JAIST), 1-1 Asahidai, Nomi, Ishikawa 923-1292, Japan}
\author{Yukiko Yamada-Takamura}
\affiliation{School of Materials Science, Japan Advanced Institute of Science and Technology (JAIST), 1-1 Asahidai, Nomi, Ishikawa 923-1292, Japan}
\author{Taisuke Ozaki}
\affiliation{Institute for Solid State Physics, The University of Tokyo, 5-1-5 Kashiwanoha, Kashiwa, Chiba 277-8581, Japan}%
\date{\today}

\begin{abstract}
Recent studies of core-level X-ray photoelectron spectroscopy (XPS) spectra of silicene on ZrB$_2$(0001) were found to be inconsistent with 
the density of states (DOS) of a planar-like structure that has been proposed as the ground state by density functional theory (DFT). To resolve the
discrepancy, a reexamination of the XPS spectra and direct theoretical access of accurate single-particle excitation energies are desired. 
By analyzing the XPS data using symmetric Voigt functions, different binding energies and its sequence of Si $2p$ orbitals can be assigned 
from previously reported ones where asymmetric pseudo-Voigt functions are adopted. Theoretically, we have adopted an approach developed very recently, which follows 
the sophisticated $\Delta$ self-consistent field ($\Delta$SCF) methods, to study the single-particle excitation of core states.
In the calculations, each single-particle energy and the renormalized core-hole charge density are calculated straightforwardly 
via two SCF calculations. By comparing the results, the theoretical core-level absolute binding energies including the splitting due to spin-orbit 
coupling are in good agreement with the observed high-resolution XPS spectra. The good agreement not only resolves the puzzling discrepancy between 
experiment and theory (DOS) but also advocates the success of DFT in describing many-body interactions of electrons at the surface.
\end{abstract}


\maketitle

\section{Introduction}
\label{sec:introduction}

Two-dimensional materials are promising for the next generation of technology and have been extensively 
studied.\cite{Castro,black,Chen,Lin,Vogt,Feng,Meng,Zhao,Fleurence} Among them, research on silicene has caught great attention since its 
successful fabrication on metallic substrates.\cite{Chen,Lin,Vogt,Feng,Meng,Zhao,Fleurence} Recent formation of isolated quasi-freestanding silicene 
showing the feature of massless Dirac fermions further remarked a considerable progress.\cite{Du} As expected, density functional theory (DFT)\cite{Hohenberg,Kohn} has provided 
substantial supports for understanding silicene.\cite{Chen,Lin,Vogt,Feng,Meng,Zhao,Fleurence,Du} In the case of silicene on ZrB$_2$(0001),\cite{Fleurence} 
the DFT-proposed ground state, the planar-like silicene\cite{Lee} as shown in Fig.~\ref{fig:energy} (a), can well reproduce the valence band structure 
measured by the angle-resolved photoelectron spectroscopy via the interpretation of Kohn-Sham eigenvalues as quasiparticle energies.\cite{Lee1,Onida,Shen} However, the core-level 
X-ray photoelectron spectroscopy (XPS)\cite{Siegbahn,Shirley,Voigt1,Voigt2,Voigt3} spectrum of silicene on ZrB$_2$(0001) was found to be consistent with the density of states 
(DOS) of Kohn-Sham orbitals of a buckled-like structure, not the planar-like one.\cite{Fleurence,Rainer} This is puzzling and implies that the underlying physics in
the core-level XPS spectrum is beyond what the DOS can offer, showing a need of theoretical access of accurate single-particle excitation energies.

The core-level excitation has been theoretically studied in the past decades.\cite{Slater,Gunnarsson,Pehlke,Blase,Car,Hellman,Cavalieri,Jeppe,Olovsson,Garcian,Bagus,Susi,Rohlfing} 
A sophisticated method to access the excitation energies is to perform the so-called $\Delta$ self-consistent field ($\Delta$SCF) calculations, 
where each excitation energy is obtained by the energy difference between the ground state and an excited state 
via two self-consistent calculations from first principles.\cite{Slater,Gunnarsson,Pehlke,Blase,Car,Hellman,Cavalieri,Jeppe,Olovsson,Garcian,Bagus,Susi} 
In the studies of single-particle excitation of core states, another accurate but time-consuming approach is to perform the quasiparticle calculation in 
the framework of diagrammatic perturbation theory within the $GW$ approximation.\cite{Rohlfing,Onida} In principle, both approaches to the core-level single-particle 
excitation energy should reach the same result that corresponds to the total energy difference between $N$ and $N-1$ electrons of a system, as given
by the pole of electronic one-particle Green's function.\cite{Rohlfing,Onida} Owing to the complicated many-body interaction of electrons, such as the effect of core-hole screening,
it is still challenging to calculate the single-particle excitation energies of core states that can directly be used to compare with the XPS experiments.

We follow the thrust of DFT that promises to give the total energy and charge density of an interacting many-electron system
to study the core-level single-particle excitation in silicene on ZrB$_{2}$(0001) for both the planar-like and buckled-like forms
by directly calculating the total energies of systems with/without a core hole using the method developed very recently.\cite{Ozaki} 
The method resembles the sophisticated $\Delta$SCF methods. Specifically, the designated core hole is introduced by adding a penalty functional through a projector.
The interaction between periodic images of the created core hole imposed by the periodic boundary condition is avoided
using an exact Coulomb-cutoff technique\cite{Payne,Ozaki} so that the calculations can reflect the nature of the ``$N-1$''-electron solid-state system. 
The calculated absolute binding energies of core states in the planar-like silicene are found to be consistent with the re-analyzed experimental data. 
The core holes are also found to be strongly dressed by other electrons. The charge density of dressed hole can be visualized in real space and is
shown to deviate from the picture of non-interacting electrons. 

The paper is organized as follows. The experimental and computational details are given in Sec.~\ref{sec:exp} and Sec.~\ref{sec:theory}, respectively.
The comparison between experimental data and theoretical results is presented in Sec.~\ref{sec:result}, where the physical quantities 
related to the single-particle excitation can also be found. In Sec.~\ref{sec:discussion} more discussions are given for addressing several
DFT-related issues. In Sec.~\ref{sec:summary} a summary is given and concludes the paper.

\section{Experimental}
\label{sec:exp}

The XPS measurements were performed using synchrotron radiation source at Photon Factory BL-13B (2015S2-008). The spectra were recorded
by a hemispherical electron energy analyzer (Scienta SES200). All the measurements were carried out at 300K in ultrahigh vacuum (UHV).
The zero binding energy was taken at the Fermi edge of a tantalum foil attached to the sample holder. Normal emission spectra
were collected. The incident photon energy was 260 eV for Si $2p$. The ZrB$_2$ film was epitaxially grown on the Si(111) substrate by chemical 
vapor deposition in UHV at JAIST. An oxide- and contamination-free epitaxial silicene monolayer was prepared on the ZrB$_2$/Si(111) 
substrate by heating the sample at 1053K for about 30 minutes in UHV. 

To analyze the contribution of each Si atom from the measured Si $2p$ spectrum (raw data), the choice of fitting functions becomes essential.
In previous studies, the raw data were fitted using asymmetric pseudo-Voigt functions for three Si atoms and found to be consistent
with the DOS of the buckled-like silicene.\cite{Fleurence,Rainer} Regarding that the Si $2p$ core levels are localized in energy away from the other states and the major energy splittings 
are the gaps between $2p_{1/2}$ and $2p_{3/2}$ orbitals, another good choice is to use the standard symmetric Voigt functions
that should be able to uncover the relevant excitation sources in Si surfaces.\cite{Voigt1,Voigt2,Voigt3} 
However, the fitted peaks using symmetric Voigt functions resemble the DOS of neither 
the buckled-like nor the planar-like silicene. Having the puzzling discrepancy between experiment and theory mentioned in Sec.~\ref{sec:introduction}, 
it is then interesting to see if the fitted peaks using symmetric Voigt functions for different photon energies could give consistent results with 
the calculated absolute binding energies of the ground-state planar-like silicene.

The fitting results using symmetric Voigt functions reveal three sets of peaks labelled as $\alpha$, $\beta$, and $\gamma$ contributed from three different Si atoms. 
The results of present study are shown in Fig.~\ref{fig:energy} (b) where each set contains two peaks corresponding to the $2p_{1/2}$ and $2p_{3/2}$ contributions. It can be observed that 
the $\alpha$ and $\gamma$ peaks have the smallest and largest binding energies, respectively, together with the in-between $\beta$ peak in each angular momentum $j$ contribution. 
In addition to the symmetric Si $2p$ peaks, shake-up components\cite{Rainer} together with a Shirley-type background have also been considered. 
The fitting results of $\alpha$, $\beta$, and $\gamma$ peaks of previously reported raw data\cite{Rainer} using symmetric Voigt functions for three different photon energies
are found to exhibit the same feature. For sake of conciseness, the corresponding $\alpha$, $\beta$, and $\gamma$ peaks are shown in the Appendix~\ref{sec:parameter}, 
where the detailed fitting parameters can also be found. 

\section{Computational detail}
\label{sec:theory}

The first-principles calculations were performed using the OpenMX code,\cite{openmx} where the addition of penalty functional 
for simulating creation of an atomic core hole and the exact Coulomb-cutoff technique for avoiding the 
undesired interaction between periodic holes were implemented.\cite{Ozaki} Although the developments allow us accessing the energy of single-particle excitation
that takes into account the effects of exchange and correlation triggered by removal of one electron, 
the approximation to the exchange-correlation functional still plays a role that could alter the results. 
In this study, the generalized gradient approximation (GGA)\cite{GGA} was adopted. The norm-conserving relativistic pseudopotentials and optimized pseudo-atomic basis functions 
were used as additional approximations to the all-electron problem.\cite{Theurich,Morrison,Ozaki1} 
In the calculations with presence of core holes, Si $2p$ core states were relaxed during the SCF iterations. 
The spin-orbit coupling was taken into account and responsible for the energy splitting between $j=1/2$ and $j=3/2$ angular momenta.
Three, two, and two optimized radial functions were allocated for the $s$, $p$, and $d$ orbitals for each Zr atom with a 
cutoff radius of 7 bohr, respectively, denoted as Zr7.0-$s3p2d2$. For Si and B atoms, Si7.0-$s2p2d1$ and B7.0-$s2p2d1$ were chosen. 
A cutoff energy of 220 Ry was used for numerical integrations and for the solution of the Poisson equation. 
The $2\times2$ $k$-point sampling was adopted for the ($4\times4$) ZrB$_2$(0001) unit cell whose in-plane lattice constant was set
to $4 \times 3.174$ \AA.\cite{Lee} The slab with five Zr layers and four B layers were studied and terminated by either the Zr layer 
or silicene. The vacuum thickness separating the slabs was set to 20 \AA. The core-level binding energies of the ($4\times4$) ZrB$_2$(0001)
were found to converge with the supercell size in the treatment for metals, where the binding energies were calculated solely in the $N$-electron system 
and would be consistent with those calculated involving the $N$ and $N-1$ electrons after reaching good convergence against the supercell size.\cite{Ozaki} 
The atomic positions were relaxed without spin-orbit coupling until the forces were less than $2\times10^{-4}$ hartree/bohr.
The optimized geometric structures were frozen for the calculations with the presence of core holes.

\section{Results}
\label{sec:result}

The structures of planar-like and buckled-like phases, obtained by the DFT calculations, are presented in Fig.~\ref{fig:energy} (a). 
The total energy of the planar-like phase is lower than that of the buckled-like phase by 272 meV per Si atom.
The buckled-like silicene resembles the freestanding silicene\cite{Cahangirov} but the lower Si atoms are pushed up while locating on top 
of Zr atoms, denoted as t-Si, in comparison with the other lower Si atoms locating above the centers of Zr triangles, 
denoted as hollow-Si (h-Si). The higher Si atoms are immune to further buckling and locate above the Zr-Zr bonds as bridging Zr atoms, 
denoted as b-Si. On the other hand, the planar-like silicene possesses a different structure, where only the t-Si is protruding 
out leaving the others at similar heights, close to a planar structure. It can be expected that these different structures should give distinguishable 
single-particle excitation energies of core states. 

\subsection{Single-particle excitation energy}
\label{sec:excitationenergy}

The physical picture of single-particle excitation can be understood by studying the difference of physical quantities between $N$
and $N-1$ (or $N+1$) particles in the many-electron system. Once a Si $2p$ electron is removed from silicene 
as a bare hole, the surrounding electrons that were correlated with the removed electron would act on, for example, screening the positive charge of 
the hole in reducing the electrostatic energy and bring the system to a new state. The new state should be described by the many-body eigenstates of 
$N-1$ electrons. To have long lifetime for the renormalized hole, the best scenario is that the new final state belongs to a single eigenstate $|m\rangle$. 
The energy of the dressed hole (DH) can be considered as $E_{DH} = E_{m}^{N-1} - E_{0}^{N}$, where $E_{0}$ denotes the ground state. 
Obviously, once the dressed hole is created, the system immediately falls into the eigenstate $|m\rangle$ with the total energy equal to $E_{0}^{N} + E_{DH}$.
After minimizing the total energy of the system with presence of the bare hole introduced by the penalty functional, the DFT-proposed state is then considered 
as the excited state $|m\rangle$. The obtained total energy can be substituted for $E_{m}^{N-1}$.

In XPS experiments, the values of binding energy $E_{B} = \mu^{N} + E_{DH}$ is obtained, where $\mu$
denotes the chemical potential. For reaching good convergence of $E_{B}$ against the supercell size in metallic systems, the binding energy can be alternatively determined by
$E_{B}^{metal} = E_{m}^{N} - E_{0}^{N}$.\cite{Ozaki} The $E_{B}^{metal}$'s of Si $2p$ orbitals calculated on the ($4\times4$) ZrB$_2$(0001) substrate 
and the convergence test of h-Si of the planar-like phase in collinear calculations as well as the observed XPS peaks are listed in Table~\ref{table:bindingenergy}. 
Note that we directly investigate the total energy difference via two SCF calculations ($\Delta$SCF) for each binding energy, where
no dynamical effects, like those described by the diagrammatic processes within the $GW$ approximation, are explicitly calculated for the screening.

\begin{table}[tbp]
\caption{Binding energies ($E_{B}^{metal}$) of Si $2p_{1/2}$ and $2p_{3/2}$ orbitals of silicene on ($4\times4$) ZrB$_2$(0001) in the
buckled-like (B-like) and planar-like (P-like) structures as well as the fitting values of XPS peaks in eV. 
The fitting values for different photon energies (PE) are given, respectively. The raw data for the PE at 130 eV, 340 eV, and 700 eV are obtained from Ref.~\onlinecite{Rainer}.
$E_{B}^{metal}$'s of P-like h-Si $2p_{z}$ orbitals against supercell sizes from collinear calculations are also listed for showing the convergence.}
\label{table:bindingenergy}%
\begin{tabular}{ccccccc}
\hline\hline
($4\times4$) &       & $2p_{1/2}$ &        &        & $2p_{3/2}$          &      \\     
             & h-Si  & b-Si    & t-Si   & h-Si   & b-Si  & t-Si    \\
B-like $E_{B}^{metal}$      & 99.33 & 99.39   & 99.14  & 98.61 & 98.70  & 98.45  \\
P-like $E_{B}^{metal}$      & 99.15 & 99.50   & 99.63  & 98.45 & 98.80  & 98.91  \\ 
experiment   & $\alpha$ & $\beta$ & $\gamma$ & $\alpha$ & $\beta$ & $\gamma$ \\ 
PE=130eV       &  99.30 & 99.55 & 99.67 & 98.69 & 98.94 & 99.06 \\         
PE=260eV       &  99.26 & 99.50 & 99.66 & 98.64 & 98.88 & 99.05 \\         
PE=340eV       &  99.34 & 99.59 & 99.77 & 98.72 & 98.97 & 99.15 \\         
PE=700eV       &  99.31 & 99.56 & 99.71 & 98.69 & 98.95 & 99.09 \\ \hline        
P-like       & ($2\times2$)   &  ($4\times4$) & ($6\times6$) & & & \\ 
h-Si         &   98.43        & 98.35         &  98.37       & & & \\ \hline\hline
\end{tabular}%
\end{table}

\begin{figure*}[tbp]
\includegraphics[width=1.75\columnwidth,clip=true,angle=0]{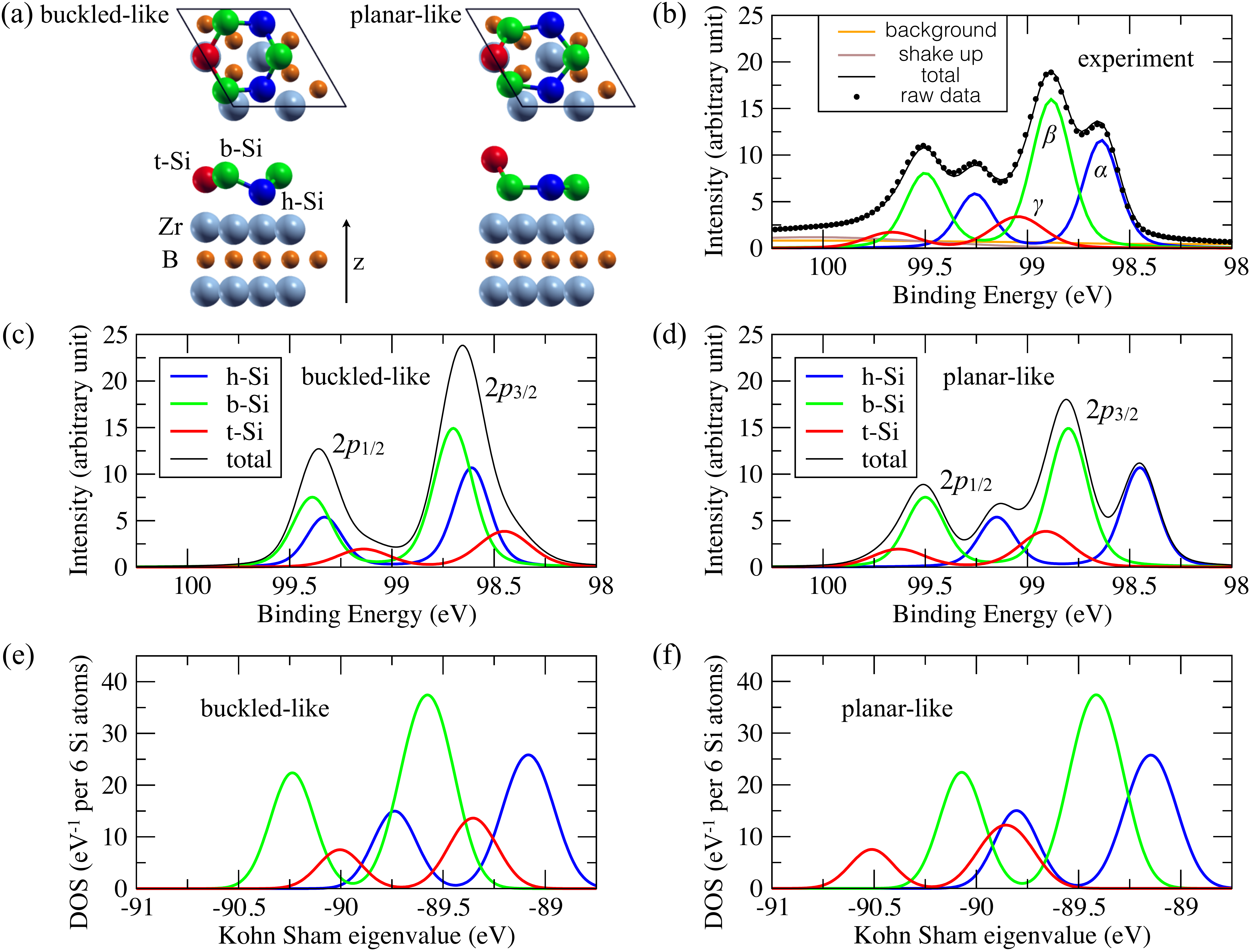}
\caption{(Color online). 
(a) Structures of buckled-like and planar-like silicene, where the t-Si (red), b-Si (green), and h-Si (blue) are indicated.
(b) Measured Si $2p$ XPS spectrum (filled dots) and fitting results by symmetric Voigt functions. The fitted $\alpha$ (blue), $\beta$ (green), and $\gamma$ (red)
peaks together with the background (orange) and shake-up (brown) contributions are presented. 
DFT-GGA binding energies are plotted by the same Voigt broadening using the energies listed in Table~\ref{table:bindingenergy}, and the area ratio
of h-Si, b-Si, and t-Si is considered as 2:3:1 for both of the (c) buckled-like and (d) planar-like silicene. 
Density of states of buckled-like and planar-like silicene with Gaussian broadening of 0.15 eV is shown in (e) and (f), respectively.
}
\label{fig:energy}
\end{figure*}

In Fig.~\ref{fig:energy} (c) and (d), the binding energies $E_{B}^{metal}$'s listed in Table~\ref{table:bindingenergy} are plotted using the same 
Voigt broadening, and the area ratio of h-Si, b-Si, and t-Si is considered as 2:3:1. Interestingly, the DFT-GGA-revealed peaks for the $2p_{1/2}$ and $2p_{3/2}$ holes 
created in h-Si, b-Si, and t-Si of the planar-like phase shown in Fig.~\ref{fig:energy} (d) agree quite well with the $\alpha$, $\beta$, and $\gamma$ peaks, respectively, 
in the absolute scale. In contrast, the t-Si $E_{B}^{metal}$ of the buckled-like phase possesses the smallest value as presented in Fig.~\ref{fig:energy} (c), which 
is inconsistent with the XPS measurements described in Sec.~\ref{sec:exp} and also in Fig.~\ref{fig:energy} (b). This finding resolves the puzzling discrepancy 
in analyzing the core-level binding energies between the experimental data and those obtained from the ground-state planar-like phase. 

Note that the previously reported XPS study compared the data to the relative eigenvalues of Kohn-Sham orbitals of the buckled-like phase, which is the information presented by the DOS.
The DOS of both phases is shown in Figs.~\ref{fig:energy} (e) and (f), respectively. The energy sequence of Kohn-Sham eigenvalues in the buckled-like phase is 
b-Si, t-Si, and h-Si, where the b-Si eigenvalue has the lowest value, while the sequence is 
t-Si, b-Si, and h-Si in the planar-like phase. These relative eigenvalues can be understood in terms of the Coulomb repulsion between orbitals. Assuming the orbitals 
distribute their charge density from each atomic center to its near neighbors, having more near neighbors would give stronger Coulomb repulsion and increase the energy. 
Each Si atom has three nearest neighbor Si atoms together with the nearest neighbor Zr atoms. Guided by the structures in Fig.~\ref{fig:energy} (a), 
the coordination numbers of t-Si, b-Si, and h-Si can be considered as four, three, and six in the buckled-like phase and three, five, and six in the planar-like phase, 
respectively, which exactly map the energy sequences in the DOS. However, these energy sequences differ from the calculated absolute binding energies.
The Kohn-Sham eigenvalues also deviate from both the measured and calculated binding energies by $\sim 9$ eV in the absolute scale.
The core-hole screening from other electrons is responsible for giving rise to both the observed XPS spectrum and 
calculated binding energies, which will be discussed in Sec.~\ref{sec:chargedensity} and Sec.~\ref{sec:discussion}.

\subsection{Charge density of dressed hole}
\label{sec:chargedensity}

We now consider the modification of charge density distribution via the creation of core hole. The modification involved in the single-particle excitation
can be defined as $n_{DH} = n_{m}^{N-1} - n_{0}^{N}$ for representing the charge density of the dressed core hole. 
Unlike the case of non-interacting electrons, the $n_{DH}$ is not purely negative in real space so that both holes and electrons can be created. 
The created electrons play an essential role in screening the holes whose major constituent is from the introduced penalty functional.
For metallic systems, $n_{DH}$ can be defined as $n_{m}^{N} - n_{0}^{N} + \delta n$, where $\delta n$ is regarded as dilute electron gas. The integration of $\delta n$ 
over the whole system should be exactly $-1$. The major response of charge density around the bare hole can then be focused on $n_{m}^{N} - n_{0}^{N}$, 
where the positive and negative values reflect the needed electrons and holes to be created in the ground state, respectively.  

\begin{figure}[tbp]
\includegraphics[width=1.00\columnwidth,clip=true,angle=0]{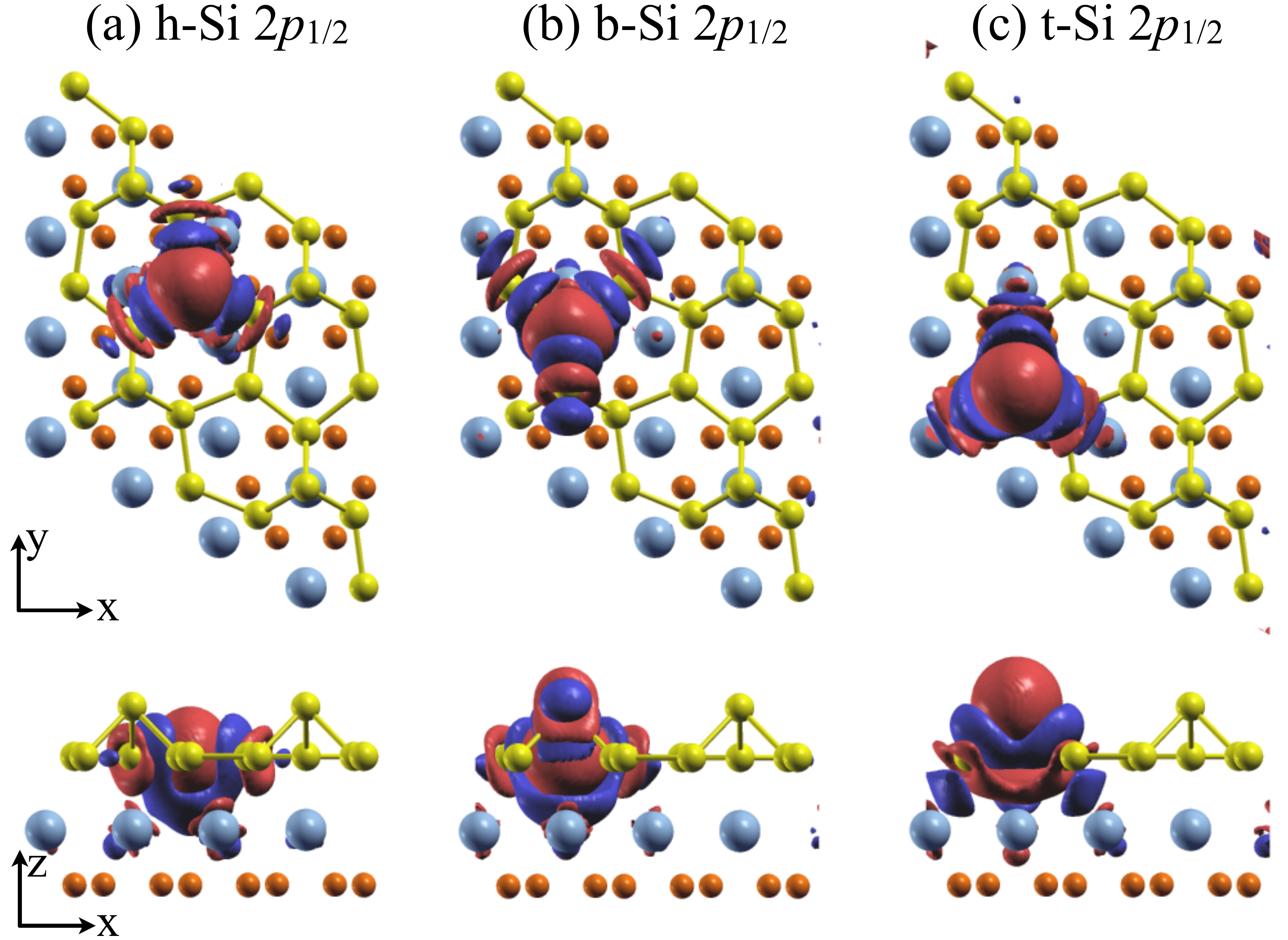}
\caption{(Color online). Charge density modification, defined as $n_{m}^{N} - n_{0}^{N}$, of planar-like silicene (yellow). The positive (red) and negative (blue) contributions 
indicate the amounts of charge density need to be added and removed from $n_{0}^{N}$, respectively. The charge-density surface is plotted at
0.001 $e/bohr^{3}$. The cases with core holes created in (a) h-Si $2p_{1/2}$, (b) b-Si $2p_{1/2}$, and (c) t-Si $2p_{1/2}$ are presented. The $2p_{3/2}$ results are similar 
to the $2p_{1/2}$ ones (not shown).}
\label{fig:chargemetal}
\end{figure}

The charge density of dressed hole can be visualized in real space. The $n_{DH}$ of planar-like phase without presenting $\delta n$ is plotted in Fig.~\ref{fig:chargemetal}. 
The positive and negative contributions corresponding to the addition and removal of electrons from the ground state, respectively, can be identified. 
The spherical shape of positive contribution surrounding the Si atom with a core hole can be seen in Fig.~\ref{fig:chargemetal}, which screens the positive charge of the core hole. 
The deviation of each dressed hole from the atomic eigenstate can also be recognized from the charge density distribution that shows large positive and negative modifications 
around the neighboring atoms, including the top-layer Zr ones. This clearly supports that the bare holes are strongly dressed
in consistency with the claimed essential role of the final-state effects in the XPS spectra.\cite{Pehlke} 
The difference of $n_{DH}$ between $2p_{1/2}$ and $2p_{3/2}$ states can reflect the effect of core-level spin-orbit coupling. 
At long distance from the core, there is no significant difference as shown by the 0.001 $e/bohr^{3}$ charge density surface in 
Fig.~\ref{fig:chargemetal}, reflecting the electrostatic screening of a positive point charge. 

In DFT the total energy is uniquely determined by the charge density,
the effect of taking into account the charge density of fully dressed holes enhances the h-Si $E_{B}^{metal}$ in the buckled-like phase and suppresses the 
t-Si $E_{B}^{metal}$ in the planar-like phase in comparison with the respective DOS. As a result, the energy sequences of the peaks in both phases are changed, 
especially in the buckled-like phase. Although the relative values of the planar-like phase in the DOS are similar to those in $E_{B}^{metal}$'s and can
be understood therein, having the accurate modification of the charge density distribution gives the well separated $2p_{1/2}$ and $2p_{3/2}$ peaks in 
the absolute values of binding energies that reach good agreement with the XPS spectrum. 

\section{Discussions}
\label{sec:discussion}

The calculated absolute binding energies of the planar-like phase are shown in good agreement with the XPS experiments. This implies 
that the total energy differences obtained in the DFT-GGA calculations are able to describe the poles of one-particle Green's function of epitaxial materials.
Recall that the ground-state total energy and charge density of a many-electron system are proven to be describable via
a set of non-interacting Kohn-Sham orbitals in the framework of DFT.\cite{Hohenberg,Kohn} The achievement has led to successful predictions of the 
ground-state properties, and DFT continues as one of the state-of-the-art methods in studying condensed matter.\cite{Jones} 
Our study of epitaxial silicene gives an example of successful applications of DFT in studying excited states of many-electron systems 
within the ground-state-type SCF calculations.

\begin{table}[tbp]
\caption{The $n_{m,0}^{N}$, defined as $\int{|n_{m}^{N}(\vec{r}) - n_{0}^{N}(\vec{r})|d\vec{r}}$, and $n_{DH}^{total}$, defined as $\int{|n_{m}^{N-1}(\vec{r}) - n_{0}^{N}(\vec{r})|d\vec{r}}$, 
of buckled-like (B-like) and planar-like (P-like) silicene on ($4\times4$) ZrB$_2$(0001) in the unit of number of electrons.}
\label{table:chargemodification}%
\begin{tabular}{ccccccc}
\hline\hline
                                     &       & $2p_{1/2}$ &        &        & $2p_{3/2}$          &      \\     
                                     & h-Si & b-Si   & t-Si   & h-Si  & b-Si   & t-Si    \\
B-like silicene ( $n_{m,0}^{N}$ )    & 3.15 & 3.09   & 3.03   & 3.18  & 3.12   & 3.08 \\
B-like silicene ( $n_{DH}^{total}$ ) & 3.60 & 3.61   & 3.57   & 3.63  & 3.64   & 3.62 \\
P-like silicene ( $n_{m,0}^{N}$ )    & 3.07 & 3.03   & 3.13   & 3.12  & 3.07   & 3.16 \\ 
P-like silicene ( $n_{DH}^{total}$ ) & 3.57 & 3.57   & 3.64   & 3.61  & 3.61   & 3.67 \\ \hline\hline
\end{tabular}%
\end{table}

The issue concerning the physical meaning of Kohn-Sham orbitals has been frequently discussed.\cite{Gunnarsson,Gonze,Stowasser,Martin} 
It is known that Kohn-Sham eigenvalues cannot be directly used to compare with the quasiparticle energies\cite{Martin,Rohlfing,Onida} that are measured in 
the photoelectron spectra.\cite{Shen,Siegbahn,Shirley,Voigt1,Voigt2,Voigt3} This makes sense since creation or annihilation of a single-particle 
eigenstate in an interacting many-body system must cause strong or weak but non-zero correlation, and the involved degrees of 
freedom in the many-body multiplets must be much larger than the number of single-particle eigenstates in general.
Nevertheless, the highest occupied orbital of a system has been proven to deliver the ionization energy.\cite{Perdew,Levy}
Each eigenvalue of Kohn-Sham orbital is also mathematically proven to connect to the derivative of the total energy with respect to its occupation.\cite{Janak} 
Our study advocates that the single-particle Kohn-Sham orbitals not only construct the ground-state charge density but are eligible for representing 
the bare hole in describing the core-level single-particle excitation. By adding a penalty functional to the designated Kohn-Sham orbital through a projector,
the new set of Kohn-Sham orbitals is able to screen the bare hole and allows accessing the excitation energy of the many-electron system in the absolute scale
even though the eigenvalues themselves could be different from the total energy differences.  

We have defined the charge density of dressed core hole $n_{DH}$ in studying the charge density modification caused by the creation of core hole.
For the metallic systems, the integration of $n_{m}^{N} - n_{0}^{N}$ over the whole system is zero with equal amounts of positive and negative values. 
The modification in charge density can then be reflected by
$n_{m,0}^{N} \equiv \int{|n_{m}^{N}(\vec{r}) - n_{0}^{N}(\vec{r})|d\vec{r}}$ integrating over the studied supercell.
As listed in Table~\ref{table:chargemodification}, $n_{m,0}^{N}$ is in the order of $3$ electrons,
which means that $\sim 1.5$ holes containing the created one and $\sim 1.5$ electrons are introduced in the ground state. 
The $\sim 1.5$ electrons are responsible for screening the created hole.
In the actual situation of metallic system of $N-1$ electrons, the electrons that screen the holes are mainly contributed from the electrons over the whole system,
that is the dilute $\delta n$ we have illustrated before. For comparison, the $n_{DH}^{total}$ defined as $\int{|n_{m}^{N-1}(\vec{r}) - n_{0}^{N}(\vec{r})|d\vec{r}}$
is also listed in Table~\ref{table:chargemodification}. Although we need a larger supercell to reach better convergence in this treatment, 
the 0.001 $e/bohr^{3}$ charge density surfaces are similar to those shown in Fig.~\ref{fig:chargemetal} and $n_{DH}^{total}$'s are also in the order of $3$ electrons. 
An estimation of leading magnitude in $n_{DH}$ can be concluded as follows: $1$ hole is from the introduced core hole, $1$ electron is in response to screening the hole, 
and the source of that electron is from the dilute electron gas distributing over the whole system, which corresponds to $1$ hole that needs to be created in the ground state.
Our study shows the importance of existence of both signs in $n_{QH}$ for delivering the accurate values of binding energies in the interacting many-electron systems.

\section{Summary}
\label{sec:summary}

Studies of creation of core holes and the induced correlations in solids are important for understanding their constituent atoms, chemical environment, 
electronic states, and various physical properties. By introducing the addition of penalty functional to simulate a core hole and the exact Coulomb-cutoff 
technique to avoid the undesired interactions between periodic images of the created core hole, we propose an approach to study the single-particle excitation 
based on the $\Delta$SCF methods. The core-level states in silicene on ZrB$_{2}$(0001) have been studied by calculating the many-body total energies
and charge density directly within the framework of DFT. From that, the addition and removal attributes of charge density of the dressed core hole that cannot be revealed by 
a single-particle eigenstate alone are demonstrated. The puzzling discrepancy between the XPS spectrum and the DOS revealed by Kohn-Sham orbitals of the ground-state 
planar-like phase has been resolved by directly comparing the absolute binding energies that are obtained from total energy differences, not 
the single-particle eigenvalues. The core holes are found to be strongly dressed by other electrons evident from the large modifications in the charge density 
involving surrounding atoms, and the energy sequence is also changed in comparison with DOS. The good agreement between theory and experiment not only resolves the 
puzzling discrepancy but also highlights the success of DFT in fingerprinting the binding energies of surface atoms including spin-orbit coupling. 
This work paves a way for future exploration of single-particle excitation of interacting electrons in various materials 
distinct from the methods based on the diagrammatic perturbation theories. \\

\begin{acknowledgments}
We are grateful for the use of supercomputers at JAIST.
This work was supported by JSPS Grant-in-Aid for Scientific Research on Innovative Area ``3D Active-Site Science'' and Priority Issue (Creation of new functional devices and high-performance materials to support next-generation industries) to be
tackled by using Post ’K’ Computer, MEXT, Japan. 
\end{acknowledgments}

\appendix

\setcounter{table}{0}
\renewcommand{\thetable}{A\arabic{table}}
\setcounter{figure}{0}
\renewcommand{\thefigure}{A\arabic{figure}}

\section{Fitting parameters}
\label{sec:parameter}

\begin{figure}[h]
\includegraphics[width=0.90\columnwidth,clip=true,angle=0]{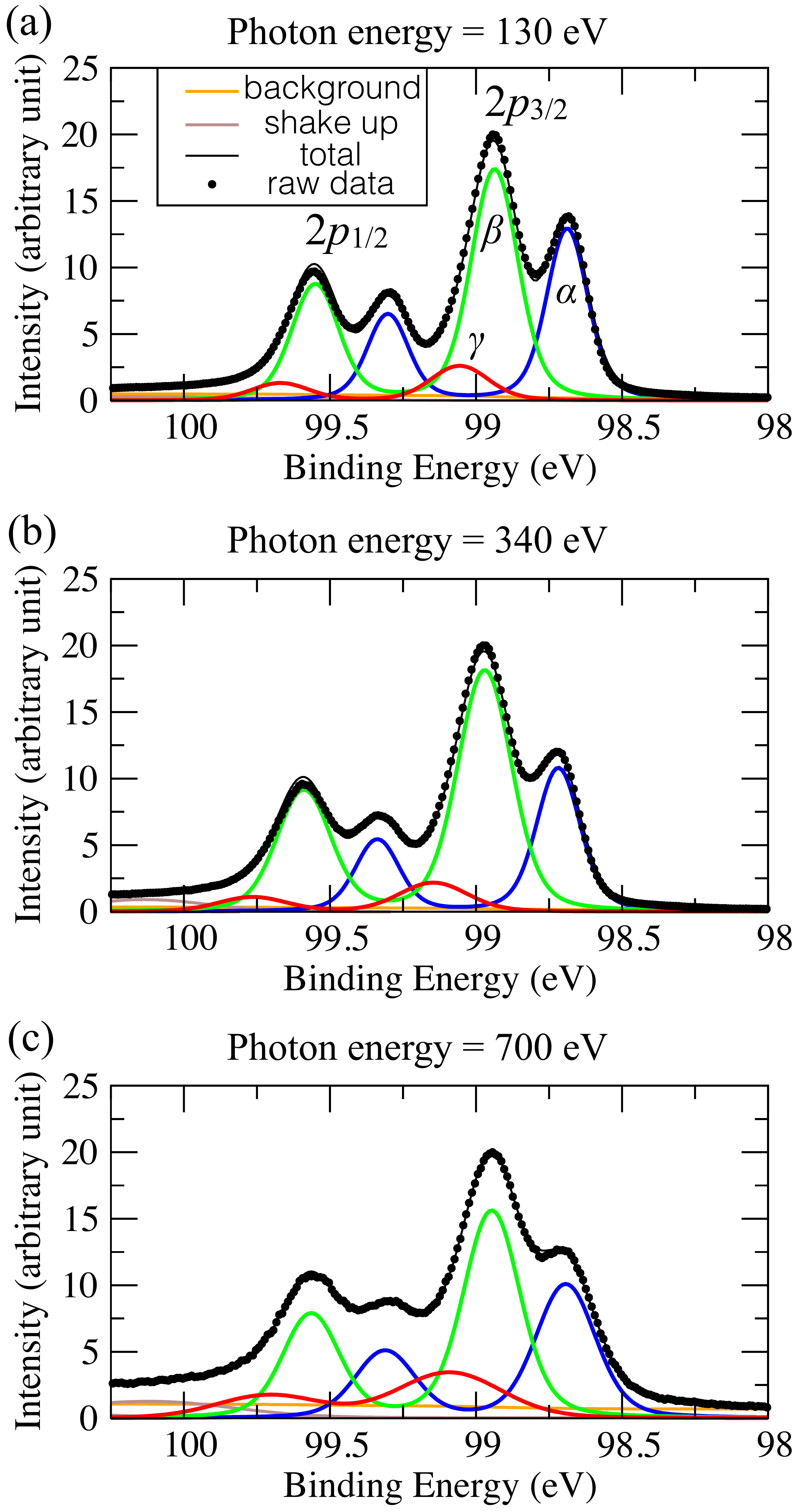}
\caption{(Color online). Measured Si $2p$ XPS spectra (filled dots) obtained from Ref.~\onlinecite{Rainer} and fitting results by symmetric Voigt functions for photon energies at (a) 130 eV, (b) 340 eV, 
and (c) 700 eV. The fitted $\alpha$ (blue), $\beta$ (green), and $\gamma$ (red) peaks together with the background (orange) and shake-up (brown) contributions 
are presented.}
\label{fig:differentenergy}
\end{figure}

The fitting parameters for different photon energies are given in Table~\ref{table:fittingparameter}.
The measured Si $2p$ spectra with the photon energies at 130 eV, 340 eV, and 700 eV are obtained from Ref.~\onlinecite{Rainer}
while the spectrum at 260 eV is from the present study. The fitted $\alpha$, $\beta$, $\gamma$, background, and shake-up peaks
together with the measured raw data are presented in Fig.~\ref{fig:differentenergy}. The larger Gaussian widths of the $\gamma$ curves 
could be the signature of the peculiar position of the pushed-up t-Si atom in the planar-like structure. 
Although the electron interference gives different patterns at different wavelengths, the intensity ratios collected
from the normal emission do not deviate from the atomic ratio 2:3:1 at different photon energies significantly.

\begin{table}[h]
\caption{Fitting parameters using symmetric Voigt functions for different photon energies (PE). The experimental raw data at 260 eV are obtained from the present study
while the raw data at 130 eV, 340 eV, and 700 eV are from Ref.~\onlinecite{Rainer}. The binding energies ($E_B$), the full width at half maximum of Lorentzian ($W_L$), and
the full width at half maximum of Gaussian ($W_G$) for each $\alpha$, $\beta$, or $\gamma$ curve are given in the unit of eV. The (area) intensity ratio (R) is also listed.
}
\label{table:fittingparameter}%
\begin{tabular}{ccccccc}
\hline\hline
 PE=130eV  & $E_B$($2p_{1/2}$) & $E_B$($2p_{3/2}$)  & $W_L$       & $W_G$       & R         \\
 $\alpha$  & 99.302         & 98.688                & 0.056       & 0.135       & 3.69        \\
 $\beta$   & 99.551         & 98.937                & 0.056       & 0.159       & 5.58        \\
 $\gamma$  & 99.670         & 99.056                & 0.056       & 0.202       & 1.00        \\ \hline
 PE=260eV  & $E_B$($2p_{1/2}$) & $E_B$($2p_{3/2}$)  & $W_L$       & $W_G$       & R         \\
 $\alpha$  & 99.257         & 98.639                & 0.062       & 0.174       & 2.47        \\
 $\beta$   & 99.501         & 98.883                & 0.062       & 0.192       & 3.65        \\
 $\gamma$  & 99.665         & 99.047                & 0.062       & 0.267       & 1.00        \\ \hline
 PE=340eV  & $E_B$($2p_{1/2}$) & $E_B$($2p_{3/2}$)  & $W_L$       & $W_G$       & R         \\
 $\alpha$  & 99.338         & 98.718                & 0.056       & 0.145       & 3.34        \\
 $\beta$   & 99.590         & 98.970                & 0.056       & 0.187       & 6.76        \\
 $\gamma$  & 99.766         & 99.146                & 0.056       & 0.245       & 1.00        \\ \hline
 PE=700eV  & $E_B$($2p_{1/2}$) & $E_B$($2p_{3/2}$)  & $W_L$       & $W_G$       & R         \\
 $\alpha$  & 99.312         & 98.693                & 0.067       & 0.204       & 1.78        \\
 $\beta$   & 99.565         & 98.946                & 0.067       & 0.191       & 2.62        \\
 $\gamma$  & 99.710         & 99.091                & 0.067       & 0.382       & 1.00        \\ \hline\hline
\end{tabular}%
\end{table}

\bibliography{refs}

\begin{thebibliography}{50}%
\makeatletter
\providecommand \@ifxundefined [1]{%
 \@ifx{#1\undefined}
}%
\providecommand \@ifnum [1]{%
 \ifnum #1\expandafter \@firstoftwo
 \else \expandafter \@secondoftwo
 \fi
}%
\providecommand \@ifx [1]{%
 \ifx #1\expandafter \@firstoftwo
 \else \expandafter \@secondoftwo
 \fi
}%
\providecommand \natexlab [1]{#1}%
\providecommand \enquote  [1]{``#1''}%
\providecommand \bibnamefont  [1]{#1}%
\providecommand \bibfnamefont [1]{#1}%
\providecommand \citenamefont [1]{#1}%
\providecommand \href@noop [0]{\@secondoftwo}%
\providecommand \href [0]{\begingroup \@sanitize@url \@href}%
\providecommand \@href[1]{\@@startlink{#1}\@@href}%
\providecommand \@@href[1]{\endgroup#1\@@endlink}%
\providecommand \@sanitize@url [0]{\catcode `\\12\catcode `\$12\catcode
  `\&12\catcode `\#12\catcode `\^12\catcode `\_12\catcode `\%12\relax}%
\providecommand \@@startlink[1]{}%
\providecommand \@@endlink[0]{}%
\providecommand \url  [0]{\begingroup\@sanitize@url \@url }%
\providecommand \@url [1]{\endgroup\@href {#1}{\urlprefix }}%
\providecommand \urlprefix  [0]{URL }%
\providecommand \Eprint [0]{\href }%
\providecommand \doibase [0]{http://dx.doi.org/}%
\providecommand \selectlanguage [0]{\@gobble}%
\providecommand \bibinfo  [0]{\@secondoftwo}%
\providecommand \bibfield  [0]{\@secondoftwo}%
\providecommand \translation [1]{[#1]}%
\providecommand \BibitemOpen [0]{}%
\providecommand \bibitemStop [0]{}%
\providecommand \bibitemNoStop [0]{.\EOS\space}%
\providecommand \EOS [0]{\spacefactor3000\relax}%
\providecommand \BibitemShut  [1]{\csname bibitem#1\endcsname}%
\let\auto@bib@innerbib\@empty
\bibitem [{\citenamefont {Castro~Neto}\ \emph {et~al.}(2009)\citenamefont
  {Castro~Neto}, \citenamefont {Guinea}, \citenamefont {Peres}, \citenamefont
  {Novoselov},\ and\ \citenamefont {Geim}}]{Castro}%
  \BibitemOpen
  \bibfield  {author} {\bibinfo {author} {\bibfnamefont {A.~H.}\ \bibnamefont
  {Castro~Neto}}, \bibinfo {author} {\bibfnamefont {F.}~\bibnamefont {Guinea}},
  \bibinfo {author} {\bibfnamefont {N.~M.~R.}\ \bibnamefont {Peres}}, \bibinfo
  {author} {\bibfnamefont {K.~S.}\ \bibnamefont {Novoselov}}, \ and\ \bibinfo
  {author} {\bibfnamefont {A.~K.}\ \bibnamefont {Geim}},\ }\href {\doibase
  10.1103/RevModPhys.81.109} {\bibfield  {journal} {\bibinfo  {journal} {Rev.
  Mod. Phys.}\ }\textbf {\bibinfo {volume} {81}},\ \bibinfo {pages} {109}
  (\bibinfo {year} {2009})}\BibitemShut {NoStop}%
\bibitem [{\citenamefont {Li}\ \emph {et~al.}(2014)\citenamefont {Li},
  \citenamefont {Yu}, \citenamefont {Ye}, \citenamefont {Ge}, \citenamefont
  {Ou}, \citenamefont {Wu}, \citenamefont {Feng}, \citenamefont {Chen},\ and\
  \citenamefont {Zhang}}]{black}%
  \BibitemOpen
  \bibfield  {author} {\bibinfo {author} {\bibfnamefont {L.}~\bibnamefont
  {Li}}, \bibinfo {author} {\bibfnamefont {Y.}~\bibnamefont {Yu}}, \bibinfo
  {author} {\bibfnamefont {G.~J.}\ \bibnamefont {Ye}}, \bibinfo {author}
  {\bibfnamefont {Q.}~\bibnamefont {Ge}}, \bibinfo {author} {\bibfnamefont
  {X.}~\bibnamefont {Ou}}, \bibinfo {author} {\bibfnamefont {H.}~\bibnamefont
  {Wu}}, \bibinfo {author} {\bibfnamefont {D.}~\bibnamefont {Feng}}, \bibinfo
  {author} {\bibfnamefont {X.~H.}\ \bibnamefont {Chen}}, \ and\ \bibinfo
  {author} {\bibfnamefont {Y.}~\bibnamefont {Zhang}},\ }\href@noop {}
  {\bibfield  {journal} {\bibinfo  {journal} {Nature Nanotech.}\ }\textbf
  {\bibinfo {volume} {9}},\ \bibinfo {pages} {372} (\bibinfo {year}
  {2014})}\BibitemShut {NoStop}%
\bibitem [{\citenamefont {Chen}\ \emph {et~al.}(2013)\citenamefont {Chen},
  \citenamefont {Li}, \citenamefont {Feng}, \citenamefont {Ding}, \citenamefont
  {Qiu}, \citenamefont {Cheng}, \citenamefont {Wu},\ and\ \citenamefont
  {Meng}}]{Chen}%
  \BibitemOpen
  \bibfield  {author} {\bibinfo {author} {\bibfnamefont {L.}~\bibnamefont
  {Chen}}, \bibinfo {author} {\bibfnamefont {H.}~\bibnamefont {Li}}, \bibinfo
  {author} {\bibfnamefont {B.}~\bibnamefont {Feng}}, \bibinfo {author}
  {\bibfnamefont {Z.}~\bibnamefont {Ding}}, \bibinfo {author} {\bibfnamefont
  {J.}~\bibnamefont {Qiu}}, \bibinfo {author} {\bibfnamefont {P.}~\bibnamefont
  {Cheng}}, \bibinfo {author} {\bibfnamefont {K.}~\bibnamefont {Wu}}, \ and\
  \bibinfo {author} {\bibfnamefont {S.}~\bibnamefont {Meng}},\ }\href {\doibase
  10.1103/PhysRevLett.110.085504} {\bibfield  {journal} {\bibinfo  {journal}
  {Phys. Rev. Lett.}\ }\textbf {\bibinfo {volume} {110}},\ \bibinfo {pages}
  {085504} (\bibinfo {year} {2013})}\BibitemShut {NoStop}%
\bibitem [{\citenamefont {Lin}\ \emph {et~al.}(2012)\citenamefont {Lin},
  \citenamefont {Arafune}, \citenamefont {Kawahara}, \citenamefont {Tsukahara},
  \citenamefont {Minamitani}, \citenamefont {Kim}, \citenamefont {Takagi},\
  and\ \citenamefont {Kawai}}]{Lin}%
  \BibitemOpen
  \bibfield  {author} {\bibinfo {author} {\bibfnamefont {C.-L.}\ \bibnamefont
  {Lin}}, \bibinfo {author} {\bibfnamefont {R.}~\bibnamefont {Arafune}},
  \bibinfo {author} {\bibfnamefont {K.}~\bibnamefont {Kawahara}}, \bibinfo
  {author} {\bibfnamefont {N.}~\bibnamefont {Tsukahara}}, \bibinfo {author}
  {\bibfnamefont {E.}~\bibnamefont {Minamitani}}, \bibinfo {author}
  {\bibfnamefont {Y.}~\bibnamefont {Kim}}, \bibinfo {author} {\bibfnamefont
  {N.}~\bibnamefont {Takagi}}, \ and\ \bibinfo {author} {\bibfnamefont
  {M.}~\bibnamefont {Kawai}},\ }\href
  {http://stacks.iop.org/1882-0786/5/i=4/a=045802} {\bibfield  {journal}
  {\bibinfo  {journal} {Appl. Phys. Express}\ }\textbf {\bibinfo {volume}
  {5}},\ \bibinfo {pages} {045802} (\bibinfo {year} {2012})}\BibitemShut
  {NoStop}%
\bibitem [{\citenamefont {Vogt}\ \emph {et~al.}(2012)\citenamefont {Vogt},
  \citenamefont {De~Padova}, \citenamefont {Quaresima}, \citenamefont {Avila},
  \citenamefont {Frantzeskakis}, \citenamefont {Asensio}, \citenamefont
  {Resta}, \citenamefont {Ealet},\ and\ \citenamefont {Le~Lay}}]{Vogt}%
  \BibitemOpen
  \bibfield  {author} {\bibinfo {author} {\bibfnamefont {P.}~\bibnamefont
  {Vogt}}, \bibinfo {author} {\bibfnamefont {P.}~\bibnamefont {De~Padova}},
  \bibinfo {author} {\bibfnamefont {C.}~\bibnamefont {Quaresima}}, \bibinfo
  {author} {\bibfnamefont {J.}~\bibnamefont {Avila}}, \bibinfo {author}
  {\bibfnamefont {E.}~\bibnamefont {Frantzeskakis}}, \bibinfo {author}
  {\bibfnamefont {M.~C.}\ \bibnamefont {Asensio}}, \bibinfo {author}
  {\bibfnamefont {A.}~\bibnamefont {Resta}}, \bibinfo {author} {\bibfnamefont
  {B.}~\bibnamefont {Ealet}}, \ and\ \bibinfo {author} {\bibfnamefont
  {G.}~\bibnamefont {Le~Lay}},\ }\href {\doibase
  10.1103/PhysRevLett.108.155501} {\bibfield  {journal} {\bibinfo  {journal}
  {Phys. Rev. Lett.}\ }\textbf {\bibinfo {volume} {108}},\ \bibinfo {pages}
  {155501} (\bibinfo {year} {2012})}\BibitemShut {NoStop}%
\bibitem [{\citenamefont {Feng}\ \emph {et~al.}(2012)\citenamefont {Feng},
  \citenamefont {Ding}, \citenamefont {Meng}, \citenamefont {Yao},
  \citenamefont {He}, \citenamefont {Cheng}, \citenamefont {Chen},\ and\
  \citenamefont {Wu}}]{Feng}%
  \BibitemOpen
  \bibfield  {author} {\bibinfo {author} {\bibfnamefont {B.}~\bibnamefont
  {Feng}}, \bibinfo {author} {\bibfnamefont {Z.}~\bibnamefont {Ding}}, \bibinfo
  {author} {\bibfnamefont {S.}~\bibnamefont {Meng}}, \bibinfo {author}
  {\bibfnamefont {Y.}~\bibnamefont {Yao}}, \bibinfo {author} {\bibfnamefont
  {X.}~\bibnamefont {He}}, \bibinfo {author} {\bibfnamefont {P.}~\bibnamefont
  {Cheng}}, \bibinfo {author} {\bibfnamefont {L.}~\bibnamefont {Chen}}, \ and\
  \bibinfo {author} {\bibfnamefont {K.}~\bibnamefont {Wu}},\ }\href@noop {}
  {\bibfield  {journal} {\bibinfo  {journal} {Nano Lett.}\ }\textbf {\bibinfo
  {volume} {12}},\ \bibinfo {pages} {3507} (\bibinfo {year}
  {2012})}\BibitemShut {NoStop}%
\bibitem [{\citenamefont {Meng}\ \emph {et~al.}(2013)\citenamefont {Meng},
  \citenamefont {Wang}, \citenamefont {Zhang}, \citenamefont {Du},
  \citenamefont {Wu}, \citenamefont {Li}, \citenamefont {Zhang}, \citenamefont
  {Li}, \citenamefont {Zhou}, \citenamefont {Hofer} \emph {et~al.}}]{Meng}%
  \BibitemOpen
  \bibfield  {author} {\bibinfo {author} {\bibfnamefont {L.}~\bibnamefont
  {Meng}}, \bibinfo {author} {\bibfnamefont {Y.}~\bibnamefont {Wang}}, \bibinfo
  {author} {\bibfnamefont {L.}~\bibnamefont {Zhang}}, \bibinfo {author}
  {\bibfnamefont {S.}~\bibnamefont {Du}}, \bibinfo {author} {\bibfnamefont
  {R.}~\bibnamefont {Wu}}, \bibinfo {author} {\bibfnamefont {L.}~\bibnamefont
  {Li}}, \bibinfo {author} {\bibfnamefont {Y.}~\bibnamefont {Zhang}}, \bibinfo
  {author} {\bibfnamefont {G.}~\bibnamefont {Li}}, \bibinfo {author}
  {\bibfnamefont {H.}~\bibnamefont {Zhou}}, \bibinfo {author} {\bibfnamefont
  {W.~A.}\ \bibnamefont {Hofer}},  \emph {et~al.},\ }\href@noop {} {\bibfield
  {journal} {\bibinfo  {journal} {Nano lett.}\ }\textbf {\bibinfo {volume}
  {13}},\ \bibinfo {pages} {685} (\bibinfo {year} {2013})}\BibitemShut
  {NoStop}%
\bibitem [{\citenamefont {Zhao}\ \emph {et~al.}(2016)\citenamefont {Zhao},
  \citenamefont {Liu}, \citenamefont {Yu}, \citenamefont {Quhe}, \citenamefont
  {Zhou}, \citenamefont {Wang}, \citenamefont {Liu}, \citenamefont {Zhong},
  \citenamefont {Han}, \citenamefont {Lu} \emph {et~al.}}]{Zhao}%
  \BibitemOpen
  \bibfield  {author} {\bibinfo {author} {\bibfnamefont {J.}~\bibnamefont
  {Zhao}}, \bibinfo {author} {\bibfnamefont {H.}~\bibnamefont {Liu}}, \bibinfo
  {author} {\bibfnamefont {Z.}~\bibnamefont {Yu}}, \bibinfo {author}
  {\bibfnamefont {R.}~\bibnamefont {Quhe}}, \bibinfo {author} {\bibfnamefont
  {S.}~\bibnamefont {Zhou}}, \bibinfo {author} {\bibfnamefont {Y.}~\bibnamefont
  {Wang}}, \bibinfo {author} {\bibfnamefont {C.~C.}\ \bibnamefont {Liu}},
  \bibinfo {author} {\bibfnamefont {H.}~\bibnamefont {Zhong}}, \bibinfo
  {author} {\bibfnamefont {N.}~\bibnamefont {Han}}, \bibinfo {author}
  {\bibfnamefont {J.}~\bibnamefont {Lu}},  \emph {et~al.},\ }\href@noop {}
  {\bibfield  {journal} {\bibinfo  {journal} {Progress in Materials Science}\
  }\textbf {\bibinfo {volume} {83}},\ \bibinfo {pages} {24} (\bibinfo {year}
  {2016})}\BibitemShut {NoStop}%
\bibitem [{\citenamefont {Fleurence}\ \emph {et~al.}(2012)\citenamefont
  {Fleurence}, \citenamefont {Friedlein}, \citenamefont {Ozaki}, \citenamefont
  {Kawai}, \citenamefont {Wang},\ and\ \citenamefont
  {Yamada-Takamura}}]{Fleurence}%
  \BibitemOpen
  \bibfield  {author} {\bibinfo {author} {\bibfnamefont {A.}~\bibnamefont
  {Fleurence}}, \bibinfo {author} {\bibfnamefont {R.}~\bibnamefont
  {Friedlein}}, \bibinfo {author} {\bibfnamefont {T.}~\bibnamefont {Ozaki}},
  \bibinfo {author} {\bibfnamefont {H.}~\bibnamefont {Kawai}}, \bibinfo
  {author} {\bibfnamefont {Y.}~\bibnamefont {Wang}}, \ and\ \bibinfo {author}
  {\bibfnamefont {Y.}~\bibnamefont {Yamada-Takamura}},\ }\href {\doibase
  10.1103/PhysRevLett.108.245501} {\bibfield  {journal} {\bibinfo  {journal}
  {Phys. Rev. Lett.}\ }\textbf {\bibinfo {volume} {108}},\ \bibinfo {pages}
  {245501} (\bibinfo {year} {2012})}\BibitemShut {NoStop}%
\bibitem [{\citenamefont {Du}\ \emph {et~al.}(2016)\citenamefont {Du},
  \citenamefont {Zhuang}, \citenamefont {Wang}, \citenamefont {Li},
  \citenamefont {Liu}, \citenamefont {Zhao}, \citenamefont {Xu}, \citenamefont
  {Feng}, \citenamefont {Chen}, \citenamefont {Wu} \emph {et~al.}}]{Du}%
  \BibitemOpen
  \bibfield  {author} {\bibinfo {author} {\bibfnamefont {Y.}~\bibnamefont
  {Du}}, \bibinfo {author} {\bibfnamefont {J.}~\bibnamefont {Zhuang}}, \bibinfo
  {author} {\bibfnamefont {J.}~\bibnamefont {Wang}}, \bibinfo {author}
  {\bibfnamefont {Z.}~\bibnamefont {Li}}, \bibinfo {author} {\bibfnamefont
  {H.}~\bibnamefont {Liu}}, \bibinfo {author} {\bibfnamefont {J.}~\bibnamefont
  {Zhao}}, \bibinfo {author} {\bibfnamefont {X.}~\bibnamefont {Xu}}, \bibinfo
  {author} {\bibfnamefont {H.}~\bibnamefont {Feng}}, \bibinfo {author}
  {\bibfnamefont {L.}~\bibnamefont {Chen}}, \bibinfo {author} {\bibfnamefont
  {K.}~\bibnamefont {Wu}},  \emph {et~al.},\ }\href@noop {} {\bibfield
  {journal} {\bibinfo  {journal} {Science Advances}\ }\textbf {\bibinfo
  {volume} {2}},\ \bibinfo {pages} {e1600067} (\bibinfo {year}
  {2016})}\BibitemShut {NoStop}%
\bibitem [{\citenamefont {Hohenberg}\ and\ \citenamefont
  {Kohn}(1964)}]{Hohenberg}%
  \BibitemOpen
  \bibfield  {author} {\bibinfo {author} {\bibfnamefont {P.}~\bibnamefont
  {Hohenberg}}\ and\ \bibinfo {author} {\bibfnamefont {W.}~\bibnamefont
  {Kohn}},\ }\href {\doibase 10.1103/PhysRev.136.B864} {\bibfield  {journal}
  {\bibinfo  {journal} {Phys. Rev.}\ }\textbf {\bibinfo {volume} {136}},\
  \bibinfo {pages} {B864} (\bibinfo {year} {1964})}\BibitemShut {NoStop}%
\bibitem [{\citenamefont {Kohn}\ and\ \citenamefont {Sham}(1965)}]{Kohn}%
  \BibitemOpen
  \bibfield  {author} {\bibinfo {author} {\bibfnamefont {W.}~\bibnamefont
  {Kohn}}\ and\ \bibinfo {author} {\bibfnamefont {L.~J.}\ \bibnamefont
  {Sham}},\ }\href {\doibase 10.1103/PhysRev.140.A1133} {\bibfield  {journal}
  {\bibinfo  {journal} {Phys. Rev.}\ }\textbf {\bibinfo {volume} {140}},\
  \bibinfo {pages} {A1133} (\bibinfo {year} {1965})}\BibitemShut {NoStop}%
\bibitem [{\citenamefont {Lee}\ \emph {et~al.}(2013)\citenamefont {Lee},
  \citenamefont {Fleurence}, \citenamefont {Friedlein}, \citenamefont
  {Yamada-Takamura},\ and\ \citenamefont {Ozaki}}]{Lee}%
  \BibitemOpen
  \bibfield  {author} {\bibinfo {author} {\bibfnamefont {C.-C.}\ \bibnamefont
  {Lee}}, \bibinfo {author} {\bibfnamefont {A.}~\bibnamefont {Fleurence}},
  \bibinfo {author} {\bibfnamefont {R.}~\bibnamefont {Friedlein}}, \bibinfo
  {author} {\bibfnamefont {Y.}~\bibnamefont {Yamada-Takamura}}, \ and\ \bibinfo
  {author} {\bibfnamefont {T.}~\bibnamefont {Ozaki}},\ }\href {\doibase
  10.1103/PhysRevB.88.165404} {\bibfield  {journal} {\bibinfo  {journal} {Phys.
  Rev. B}\ }\textbf {\bibinfo {volume} {88}},\ \bibinfo {pages} {165404}
  (\bibinfo {year} {2013})}\BibitemShut {NoStop}%
\bibitem [{\citenamefont {Lee}\ \emph {et~al.}(2014)\citenamefont {Lee},
  \citenamefont {Fleurence}, \citenamefont {Yamada-Takamura}, \citenamefont
  {Ozaki},\ and\ \citenamefont {Friedlein}}]{Lee1}%
  \BibitemOpen
  \bibfield  {author} {\bibinfo {author} {\bibfnamefont {C.-C.}\ \bibnamefont
  {Lee}}, \bibinfo {author} {\bibfnamefont {A.}~\bibnamefont {Fleurence}},
  \bibinfo {author} {\bibfnamefont {Y.}~\bibnamefont {Yamada-Takamura}},
  \bibinfo {author} {\bibfnamefont {T.}~\bibnamefont {Ozaki}}, \ and\ \bibinfo
  {author} {\bibfnamefont {R.}~\bibnamefont {Friedlein}},\ }\href {\doibase
  10.1103/PhysRevB.90.075422} {\bibfield  {journal} {\bibinfo  {journal} {Phys.
  Rev. B}\ }\textbf {\bibinfo {volume} {90}},\ \bibinfo {pages} {075422}
  (\bibinfo {year} {2014})}\BibitemShut {NoStop}%
\bibitem [{\citenamefont {Onida}\ \emph {et~al.}(2002)\citenamefont {Onida},
  \citenamefont {Reining},\ and\ \citenamefont {Rubio}}]{Onida}%
  \BibitemOpen
  \bibfield  {author} {\bibinfo {author} {\bibfnamefont {G.}~\bibnamefont
  {Onida}}, \bibinfo {author} {\bibfnamefont {L.}~\bibnamefont {Reining}}, \
  and\ \bibinfo {author} {\bibfnamefont {A.}~\bibnamefont {Rubio}},\ }\href
  {\doibase 10.1103/RevModPhys.74.601} {\bibfield  {journal} {\bibinfo
  {journal} {Rev. Mod. Phys.}\ }\textbf {\bibinfo {volume} {74}},\ \bibinfo
  {pages} {601} (\bibinfo {year} {2002})}\BibitemShut {NoStop}%
\bibitem [{\citenamefont {Damascelli}\ \emph {et~al.}(2003)\citenamefont
  {Damascelli}, \citenamefont {Hussain},\ and\ \citenamefont {Shen}}]{Shen}%
  \BibitemOpen
  \bibfield  {author} {\bibinfo {author} {\bibfnamefont {A.}~\bibnamefont
  {Damascelli}}, \bibinfo {author} {\bibfnamefont {Z.}~\bibnamefont {Hussain}},
  \ and\ \bibinfo {author} {\bibfnamefont {Z.-X.}\ \bibnamefont {Shen}},\
  }\href {\doibase 10.1103/RevModPhys.75.473} {\bibfield  {journal} {\bibinfo
  {journal} {Rev. Mod. Phys.}\ }\textbf {\bibinfo {volume} {75}},\ \bibinfo
  {pages} {473} (\bibinfo {year} {2003})}\BibitemShut {NoStop}%
\bibitem [{\citenamefont {Siegbahn}\ \emph {et~al.}(1967)\citenamefont
  {Siegbahn} \emph {et~al.}}]{Siegbahn}%
  \BibitemOpen
  \bibfield  {author} {\bibinfo {author} {\bibfnamefont {K.}~\bibnamefont
  {Siegbahn}} \emph {et~al.},\ }\href@noop {} {\emph {\bibinfo {title} {ESCA;
  atomic, molecular and solid state structure studied by means of electron
  spectroscopy}}},\ Vol.~\bibinfo {volume} {20}\ (\bibinfo  {publisher}
  {Almqvist \& Wiksells},\ \bibinfo {year} {1967})\BibitemShut {NoStop}%
\bibitem [{\citenamefont {Fadley}\ and\ \citenamefont
  {Shirley}(1968)}]{Shirley}%
  \BibitemOpen
  \bibfield  {author} {\bibinfo {author} {\bibfnamefont {C.~S.}\ \bibnamefont
  {Fadley}}\ and\ \bibinfo {author} {\bibfnamefont {D.~A.}\ \bibnamefont
  {Shirley}},\ }\href {\doibase 10.1103/PhysRevLett.21.980} {\bibfield
  {journal} {\bibinfo  {journal} {Phys. Rev. Lett.}\ }\textbf {\bibinfo
  {volume} {21}},\ \bibinfo {pages} {980} (\bibinfo {year} {1968})}\BibitemShut
  {NoStop}%
\bibitem [{\citenamefont {Landemark}\ \emph {et~al.}(1992)\citenamefont
  {Landemark}, \citenamefont {Karlsson}, \citenamefont {Chao},\ and\
  \citenamefont {Uhrberg}}]{Voigt1}%
  \BibitemOpen
  \bibfield  {author} {\bibinfo {author} {\bibfnamefont {E.}~\bibnamefont
  {Landemark}}, \bibinfo {author} {\bibfnamefont {C.~J.}\ \bibnamefont
  {Karlsson}}, \bibinfo {author} {\bibfnamefont {Y.-C.}\ \bibnamefont {Chao}},
  \ and\ \bibinfo {author} {\bibfnamefont {R.~I.~G.}\ \bibnamefont {Uhrberg}},\
  }\href {\doibase 10.1103/PhysRevLett.69.1588} {\bibfield  {journal} {\bibinfo
   {journal} {Phys. Rev. Lett.}\ }\textbf {\bibinfo {volume} {69}},\ \bibinfo
  {pages} {1588} (\bibinfo {year} {1992})}\BibitemShut {NoStop}%
\bibitem [{\citenamefont {Yamashita}\ \emph {et~al.}(2001)\citenamefont
  {Yamashita}, \citenamefont {Nagao}, \citenamefont {Machida}, \citenamefont
  {Hamaguchi}, \citenamefont {Yasui}, \citenamefont {Mukai},\ and\
  \citenamefont {Yoshinobu}}]{Voigt2}%
  \BibitemOpen
  \bibfield  {author} {\bibinfo {author} {\bibfnamefont {Y.}~\bibnamefont
  {Yamashita}}, \bibinfo {author} {\bibfnamefont {M.}~\bibnamefont {Nagao}},
  \bibinfo {author} {\bibfnamefont {S.}~\bibnamefont {Machida}}, \bibinfo
  {author} {\bibfnamefont {K.}~\bibnamefont {Hamaguchi}}, \bibinfo {author}
  {\bibfnamefont {F.}~\bibnamefont {Yasui}}, \bibinfo {author} {\bibfnamefont
  {K.}~\bibnamefont {Mukai}}, \ and\ \bibinfo {author} {\bibfnamefont
  {J.}~\bibnamefont {Yoshinobu}},\ }\href@noop {} {\bibfield  {journal}
  {\bibinfo  {journal} {J. Electron Spectrosc. Relat. Phenom.}\ }\textbf
  {\bibinfo {volume} {114-116}},\ \bibinfo {pages} {389} (\bibinfo {year}
  {2001})}\BibitemShut {NoStop}%
\bibitem [{\citenamefont {Chung}\ \emph {et~al.}(2002)\citenamefont {Chung},
  \citenamefont {Kim}, \citenamefont {Whang},\ and\ \citenamefont
  {Yeom}}]{Voigt3}%
  \BibitemOpen
  \bibfield  {author} {\bibinfo {author} {\bibfnamefont {Y.~D.}\ \bibnamefont
  {Chung}}, \bibinfo {author} {\bibfnamefont {J.~W.}\ \bibnamefont {Kim}},
  \bibinfo {author} {\bibfnamefont {C.~N.}\ \bibnamefont {Whang}}, \ and\
  \bibinfo {author} {\bibfnamefont {H.~W.}\ \bibnamefont {Yeom}},\ }\href
  {\doibase 10.1103/PhysRevB.65.155310} {\bibfield  {journal} {\bibinfo
  {journal} {Phys. Rev. B}\ }\textbf {\bibinfo {volume} {65}},\ \bibinfo
  {pages} {155310} (\bibinfo {year} {2002})}\BibitemShut {NoStop}%
\bibitem [{\citenamefont {Friedlein}\ \emph {et~al.}(2014)\citenamefont
  {Friedlein}, \citenamefont {Fleurence}, \citenamefont {Aoyagi}, \citenamefont
  {de~Jong}, \citenamefont {Van~Bui}, \citenamefont {Wiggers}, \citenamefont
  {Yoshimoto}, \citenamefont {Koitaya}, \citenamefont {Shimizu}, \citenamefont
  {Noritake} \emph {et~al.}}]{Rainer}%
  \BibitemOpen
  \bibfield  {author} {\bibinfo {author} {\bibfnamefont {R.}~\bibnamefont
  {Friedlein}}, \bibinfo {author} {\bibfnamefont {A.}~\bibnamefont
  {Fleurence}}, \bibinfo {author} {\bibfnamefont {K.}~\bibnamefont {Aoyagi}},
  \bibinfo {author} {\bibfnamefont {M.}~\bibnamefont {de~Jong}}, \bibinfo
  {author} {\bibfnamefont {H.}~\bibnamefont {Van~Bui}}, \bibinfo {author}
  {\bibfnamefont {F.}~\bibnamefont {Wiggers}}, \bibinfo {author} {\bibfnamefont
  {S.}~\bibnamefont {Yoshimoto}}, \bibinfo {author} {\bibfnamefont
  {T.}~\bibnamefont {Koitaya}}, \bibinfo {author} {\bibfnamefont
  {S.}~\bibnamefont {Shimizu}}, \bibinfo {author} {\bibfnamefont
  {H.}~\bibnamefont {Noritake}},  \emph {et~al.},\ }\href@noop {} {\bibfield
  {journal} {\bibinfo  {journal} {J. Chem. Phys.}\ }\textbf {\bibinfo {volume}
  {140}},\ \bibinfo {pages} {184704} (\bibinfo {year} {2014})}\BibitemShut
  {NoStop}%
\bibitem [{\citenamefont {Slater}(1974)}]{Slater}%
  \BibitemOpen
  \bibfield  {author} {\bibinfo {author} {\bibfnamefont {J.~C.}\ \bibnamefont
  {Slater}},\ }\href@noop {} {\emph {\bibinfo {title} {The self-consistent
  field for molecules and solids}}},\ Vol.~\bibinfo {volume} {4}\ (\bibinfo
  {publisher} {McGraw-Hill New York},\ \bibinfo {year} {1974})\BibitemShut
  {NoStop}%
\bibitem [{\citenamefont {Jones}\ and\ \citenamefont
  {Gunnarsson}(1989)}]{Gunnarsson}%
  \BibitemOpen
  \bibfield  {author} {\bibinfo {author} {\bibfnamefont {R.~O.}\ \bibnamefont
  {Jones}}\ and\ \bibinfo {author} {\bibfnamefont {O.}~\bibnamefont
  {Gunnarsson}},\ }\href {\doibase 10.1103/RevModPhys.61.689} {\bibfield
  {journal} {\bibinfo  {journal} {Rev. Mod. Phys.}\ }\textbf {\bibinfo {volume}
  {61}},\ \bibinfo {pages} {689} (\bibinfo {year} {1989})}\BibitemShut
  {NoStop}%
\bibitem [{\citenamefont {Pehlke}\ and\ \citenamefont
  {Scheffler}(1993)}]{Pehlke}%
  \BibitemOpen
  \bibfield  {author} {\bibinfo {author} {\bibfnamefont {E.}~\bibnamefont
  {Pehlke}}\ and\ \bibinfo {author} {\bibfnamefont {M.}~\bibnamefont
  {Scheffler}},\ }\href {\doibase 10.1103/PhysRevLett.71.2338} {\bibfield
  {journal} {\bibinfo  {journal} {Phys. Rev. Lett.}\ }\textbf {\bibinfo
  {volume} {71}},\ \bibinfo {pages} {2338} (\bibinfo {year}
  {1993})}\BibitemShut {NoStop}%
\bibitem [{\citenamefont {Blase}\ \emph {et~al.}(1994)\citenamefont {Blase},
  \citenamefont {da~Silva}, \citenamefont {Zhu},\ and\ \citenamefont
  {Louie}}]{Blase}%
  \BibitemOpen
  \bibfield  {author} {\bibinfo {author} {\bibfnamefont {X.}~\bibnamefont
  {Blase}}, \bibinfo {author} {\bibfnamefont {A.~J.~R.}\ \bibnamefont
  {da~Silva}}, \bibinfo {author} {\bibfnamefont {X.}~\bibnamefont {Zhu}}, \
  and\ \bibinfo {author} {\bibfnamefont {S.~G.}\ \bibnamefont {Louie}},\ }\href
  {\doibase 10.1103/PhysRevB.50.8102} {\bibfield  {journal} {\bibinfo
  {journal} {Phys. Rev. B}\ }\textbf {\bibinfo {volume} {50}},\ \bibinfo
  {pages} {8102} (\bibinfo {year} {1994})}\BibitemShut {NoStop}%
\bibitem [{\citenamefont {Pasquarello}\ \emph {et~al.}(1996)\citenamefont
  {Pasquarello}, \citenamefont {Hybertsen},\ and\ \citenamefont {Car}}]{Car}%
  \BibitemOpen
  \bibfield  {author} {\bibinfo {author} {\bibfnamefont {A.}~\bibnamefont
  {Pasquarello}}, \bibinfo {author} {\bibfnamefont {M.~S.}\ \bibnamefont
  {Hybertsen}}, \ and\ \bibinfo {author} {\bibfnamefont {R.}~\bibnamefont
  {Car}},\ }\href {\doibase 10.1103/PhysRevB.53.10942} {\bibfield  {journal}
  {\bibinfo  {journal} {Phys. Rev. B}\ }\textbf {\bibinfo {volume} {53}},\
  \bibinfo {pages} {10942} (\bibinfo {year} {1996})}\BibitemShut {NoStop}%
\bibitem [{\citenamefont {Hellman}\ \emph {et~al.}(2004)\citenamefont
  {Hellman}, \citenamefont {Razaznejad},\ and\ \citenamefont
  {Lundqvist}}]{Hellman}%
  \BibitemOpen
  \bibfield  {author} {\bibinfo {author} {\bibfnamefont {A.}~\bibnamefont
  {Hellman}}, \bibinfo {author} {\bibfnamefont {B.}~\bibnamefont {Razaznejad}},
  \ and\ \bibinfo {author} {\bibfnamefont {B.~I.}\ \bibnamefont {Lundqvist}},\
  }\href@noop {} {\bibfield  {journal} {\bibinfo  {journal} {The Journal of
  chemical physics}\ }\textbf {\bibinfo {volume} {120}},\ \bibinfo {pages}
  {4593} (\bibinfo {year} {2004})}\BibitemShut {NoStop}%
\bibitem [{\citenamefont {Cavalieri}\ \emph {et~al.}(2007)\citenamefont
  {Cavalieri}, \citenamefont {M{\"u}ller}, \citenamefont {Uphues},
  \citenamefont {Yakovlev}, \citenamefont {Baltu{\v{s}}ka}, \citenamefont
  {Horvath}, \citenamefont {Schmidt}, \citenamefont {Bl{\"u}mel}, \citenamefont
  {Holzwarth}, \citenamefont {Hendel} \emph {et~al.}}]{Cavalieri}%
  \BibitemOpen
  \bibfield  {author} {\bibinfo {author} {\bibfnamefont {A.~L.}\ \bibnamefont
  {Cavalieri}}, \bibinfo {author} {\bibfnamefont {N.}~\bibnamefont
  {M{\"u}ller}}, \bibinfo {author} {\bibfnamefont {T.}~\bibnamefont {Uphues}},
  \bibinfo {author} {\bibfnamefont {V.~S.}\ \bibnamefont {Yakovlev}}, \bibinfo
  {author} {\bibfnamefont {A.}~\bibnamefont {Baltu{\v{s}}ka}}, \bibinfo
  {author} {\bibfnamefont {B.}~\bibnamefont {Horvath}}, \bibinfo {author}
  {\bibfnamefont {B.}~\bibnamefont {Schmidt}}, \bibinfo {author} {\bibfnamefont
  {L.}~\bibnamefont {Bl{\"u}mel}}, \bibinfo {author} {\bibfnamefont
  {R.}~\bibnamefont {Holzwarth}}, \bibinfo {author} {\bibfnamefont
  {S.}~\bibnamefont {Hendel}},  \emph {et~al.},\ }\href@noop {} {\bibfield
  {journal} {\bibinfo  {journal} {Nature}\ }\textbf {\bibinfo {volume} {449}},\
  \bibinfo {pages} {1029} (\bibinfo {year} {2007})}\BibitemShut {NoStop}%
\bibitem [{\citenamefont {Gavnholt}\ \emph {et~al.}(2008)\citenamefont
  {Gavnholt}, \citenamefont {Olsen}, \citenamefont {Engelund},\ and\
  \citenamefont {Schi\o{}tz}}]{Jeppe}%
  \BibitemOpen
  \bibfield  {author} {\bibinfo {author} {\bibfnamefont {J.}~\bibnamefont
  {Gavnholt}}, \bibinfo {author} {\bibfnamefont {T.}~\bibnamefont {Olsen}},
  \bibinfo {author} {\bibfnamefont {M.}~\bibnamefont {Engelund}}, \ and\
  \bibinfo {author} {\bibfnamefont {J.}~\bibnamefont {Schi\o{}tz}},\ }\href
  {\doibase 10.1103/PhysRevB.78.075441} {\bibfield  {journal} {\bibinfo
  {journal} {Phys. Rev. B}\ }\textbf {\bibinfo {volume} {78}},\ \bibinfo
  {pages} {075441} (\bibinfo {year} {2008})}\BibitemShut {NoStop}%
\bibitem [{\citenamefont {Olovsson}\ \emph {et~al.}(2010)\citenamefont
  {Olovsson}, \citenamefont {Marten}, \citenamefont {Holmstr{\"o}m},
  \citenamefont {Johansson},\ and\ \citenamefont {Abrikosov}}]{Olovsson}%
  \BibitemOpen
  \bibfield  {author} {\bibinfo {author} {\bibfnamefont {W.}~\bibnamefont
  {Olovsson}}, \bibinfo {author} {\bibfnamefont {T.}~\bibnamefont {Marten}},
  \bibinfo {author} {\bibfnamefont {E.}~\bibnamefont {Holmstr{\"o}m}}, \bibinfo
  {author} {\bibfnamefont {B.}~\bibnamefont {Johansson}}, \ and\ \bibinfo
  {author} {\bibfnamefont {I.~A.}\ \bibnamefont {Abrikosov}},\ }\href@noop {}
  {\bibfield  {journal} {\bibinfo  {journal} {J. Electron Spectrosc. Relat.
  Phenom.}\ }\textbf {\bibinfo {volume} {178}},\ \bibinfo {pages} {88}
  (\bibinfo {year} {2010})}\BibitemShut {NoStop}%
\bibitem [{\citenamefont {Garc{\'\i}a-Gil}\ \emph {et~al.}(2012)\citenamefont
  {Garc{\'\i}a-Gil}, \citenamefont {Garc{\'\i}a},\ and\ \citenamefont
  {Ordej{\'o}n}}]{Garcian}%
  \BibitemOpen
  \bibfield  {author} {\bibinfo {author} {\bibfnamefont {S.}~\bibnamefont
  {Garc{\'\i}a-Gil}}, \bibinfo {author} {\bibfnamefont {A.}~\bibnamefont
  {Garc{\'\i}a}}, \ and\ \bibinfo {author} {\bibfnamefont {P.}~\bibnamefont
  {Ordej{\'o}n}},\ }\href@noop {} {\bibfield  {journal} {\bibinfo  {journal}
  {Eur. Phys. J. B}\ }\textbf {\bibinfo {volume} {85}},\ \bibinfo {pages} {239}
  (\bibinfo {year} {2012})}\BibitemShut {NoStop}%
\bibitem [{\citenamefont {Bagus}\ \emph {et~al.}(2013)\citenamefont {Bagus},
  \citenamefont {Ilton},\ and\ \citenamefont {Nelin}}]{Bagus}%
  \BibitemOpen
  \bibfield  {author} {\bibinfo {author} {\bibfnamefont {P.~S.}\ \bibnamefont
  {Bagus}}, \bibinfo {author} {\bibfnamefont {E.~S.}\ \bibnamefont {Ilton}}, \
  and\ \bibinfo {author} {\bibfnamefont {C.~J.}\ \bibnamefont {Nelin}},\
  }\href@noop {} {\bibfield  {journal} {\bibinfo  {journal} {Surface Science
  Reports}\ }\textbf {\bibinfo {volume} {68}},\ \bibinfo {pages} {273}
  (\bibinfo {year} {2013})}\BibitemShut {NoStop}%
\bibitem [{\citenamefont {Susi}\ \emph {et~al.}(2015)\citenamefont {Susi},
  \citenamefont {Mowbray}, \citenamefont {Ljungberg},\ and\ \citenamefont
  {Ayala}}]{Susi}%
  \BibitemOpen
  \bibfield  {author} {\bibinfo {author} {\bibfnamefont {T.}~\bibnamefont
  {Susi}}, \bibinfo {author} {\bibfnamefont {D.~J.}\ \bibnamefont {Mowbray}},
  \bibinfo {author} {\bibfnamefont {M.~P.}\ \bibnamefont {Ljungberg}}, \ and\
  \bibinfo {author} {\bibfnamefont {P.}~\bibnamefont {Ayala}},\ }\href
  {\doibase 10.1103/PhysRevB.91.081401} {\bibfield  {journal} {\bibinfo
  {journal} {Phys. Rev. B}\ }\textbf {\bibinfo {volume} {91}},\ \bibinfo
  {pages} {081401} (\bibinfo {year} {2015})}\BibitemShut {NoStop}%
\bibitem [{\citenamefont {Rohlfing}\ \emph {et~al.}(1997)\citenamefont
  {Rohlfing}, \citenamefont {Kr{\"u}ger},\ and\ \citenamefont
  {Pollmann}}]{Rohlfing}%
  \BibitemOpen
  \bibfield  {author} {\bibinfo {author} {\bibfnamefont {M.}~\bibnamefont
  {Rohlfing}}, \bibinfo {author} {\bibfnamefont {P.}~\bibnamefont
  {Kr{\"u}ger}}, \ and\ \bibinfo {author} {\bibfnamefont {J.}~\bibnamefont
  {Pollmann}},\ }\href@noop {} {\bibfield  {journal} {\bibinfo  {journal}
  {Phys. Rev. B}\ }\textbf {\bibinfo {volume} {56}},\ \bibinfo {pages} {2191}
  (\bibinfo {year} {1997})}\BibitemShut {NoStop}%
\bibitem [{\citenamefont {Ozaki}\ and\ \citenamefont {Lee}(2017)}]{Ozaki}%
  \BibitemOpen
  \bibfield  {author} {\bibinfo {author} {\bibfnamefont {T.}~\bibnamefont
  {Ozaki}}\ and\ \bibinfo {author} {\bibfnamefont {C.-C.}\ \bibnamefont
  {Lee}},\ }\href {\doibase 10.1103/PhysRevLett.118.026401} {\bibfield
  {journal} {\bibinfo  {journal} {Phys. Rev. Lett.}\ }\textbf {\bibinfo
  {volume} {118}},\ \bibinfo {pages} {026401} (\bibinfo {year}
  {2017})}\BibitemShut {NoStop}%
\bibitem [{\citenamefont {Jarvis}\ \emph {et~al.}(1997)\citenamefont {Jarvis},
  \citenamefont {White}, \citenamefont {Godby},\ and\ \citenamefont
  {Payne}}]{Payne}%
  \BibitemOpen
  \bibfield  {author} {\bibinfo {author} {\bibfnamefont {M.~R.}\ \bibnamefont
  {Jarvis}}, \bibinfo {author} {\bibfnamefont {I.~D.}\ \bibnamefont {White}},
  \bibinfo {author} {\bibfnamefont {R.~W.}\ \bibnamefont {Godby}}, \ and\
  \bibinfo {author} {\bibfnamefont {M.~C.}\ \bibnamefont {Payne}},\ }\href
  {\doibase 10.1103/PhysRevB.56.14972} {\bibfield  {journal} {\bibinfo
  {journal} {Phys. Rev. B}\ }\textbf {\bibinfo {volume} {56}},\ \bibinfo
  {pages} {14972} (\bibinfo {year} {1997})}\BibitemShut {NoStop}%
\bibitem [{ope()}]{openmx}%
  \BibitemOpen
  \href@noop {} {}\bibinfo {note} {The code, OpenMX, pseudo-atomic basis
  functions, and pseudopotentials are available on a web site
  (http://www.openmx-square.org/)}\BibitemShut {NoStop}%
\bibitem [{\citenamefont {Perdew}\ \emph {et~al.}(1996)\citenamefont {Perdew},
  \citenamefont {Burke},\ and\ \citenamefont {Ernzerhof}}]{GGA}%
  \BibitemOpen
  \bibfield  {author} {\bibinfo {author} {\bibfnamefont {J.~P.}\ \bibnamefont
  {Perdew}}, \bibinfo {author} {\bibfnamefont {K.}~\bibnamefont {Burke}}, \
  and\ \bibinfo {author} {\bibfnamefont {M.}~\bibnamefont {Ernzerhof}},\ }\href
  {\doibase 10.1103/PhysRevLett.77.3865} {\bibfield  {journal} {\bibinfo
  {journal} {Phys. Rev. Lett.}\ }\textbf {\bibinfo {volume} {77}},\ \bibinfo
  {pages} {3865} (\bibinfo {year} {1996})}\BibitemShut {NoStop}%
\bibitem [{\citenamefont {Theurich}\ and\ \citenamefont
  {Hill}(2001)}]{Theurich}%
  \BibitemOpen
  \bibfield  {author} {\bibinfo {author} {\bibfnamefont {G.}~\bibnamefont
  {Theurich}}\ and\ \bibinfo {author} {\bibfnamefont {N.~A.}\ \bibnamefont
  {Hill}},\ }\href {\doibase 10.1103/PhysRevB.64.073106} {\bibfield  {journal}
  {\bibinfo  {journal} {Phys. Rev. B}\ }\textbf {\bibinfo {volume} {64}},\
  \bibinfo {pages} {073106} (\bibinfo {year} {2001})}\BibitemShut {NoStop}%
\bibitem [{\citenamefont {Morrison}\ \emph {et~al.}(1993)\citenamefont
  {Morrison}, \citenamefont {Bylander},\ and\ \citenamefont
  {Kleinman}}]{Morrison}%
  \BibitemOpen
  \bibfield  {author} {\bibinfo {author} {\bibfnamefont {I.}~\bibnamefont
  {Morrison}}, \bibinfo {author} {\bibfnamefont {D.~M.}\ \bibnamefont
  {Bylander}}, \ and\ \bibinfo {author} {\bibfnamefont {L.}~\bibnamefont
  {Kleinman}},\ }\href {\doibase 10.1103/PhysRevB.47.6728} {\bibfield
  {journal} {\bibinfo  {journal} {Phys. Rev. B}\ }\textbf {\bibinfo {volume}
  {47}},\ \bibinfo {pages} {6728} (\bibinfo {year} {1993})}\BibitemShut
  {NoStop}%
\bibitem [{\citenamefont {Ozaki}(2003)}]{Ozaki1}%
  \BibitemOpen
  \bibfield  {author} {\bibinfo {author} {\bibfnamefont {T.}~\bibnamefont
  {Ozaki}},\ }\href {\doibase 10.1103/PhysRevB.67.155108} {\bibfield  {journal}
  {\bibinfo  {journal} {Phys. Rev. B}\ }\textbf {\bibinfo {volume} {67}},\
  \bibinfo {pages} {155108} (\bibinfo {year} {2003})}\BibitemShut {NoStop}%
\bibitem [{\citenamefont {Cahangirov}\ \emph {et~al.}(2009)\citenamefont
  {Cahangirov}, \citenamefont {Topsakal}, \citenamefont {Akt\"urk},
  \citenamefont {\ifmmode~\mbox{\c{S}}\else \c{S}\fi{}ahin},\ and\
  \citenamefont {Ciraci}}]{Cahangirov}%
  \BibitemOpen
  \bibfield  {author} {\bibinfo {author} {\bibfnamefont {S.}~\bibnamefont
  {Cahangirov}}, \bibinfo {author} {\bibfnamefont {M.}~\bibnamefont
  {Topsakal}}, \bibinfo {author} {\bibfnamefont {E.}~\bibnamefont {Akt\"urk}},
  \bibinfo {author} {\bibfnamefont {H.}~\bibnamefont
  {\ifmmode~\mbox{\c{S}}\else \c{S}\fi{}ahin}}, \ and\ \bibinfo {author}
  {\bibfnamefont {S.}~\bibnamefont {Ciraci}},\ }\href {\doibase
  10.1103/PhysRevLett.102.236804} {\bibfield  {journal} {\bibinfo  {journal}
  {Phys. Rev. Lett.}\ }\textbf {\bibinfo {volume} {102}},\ \bibinfo {pages}
  {236804} (\bibinfo {year} {2009})}\BibitemShut {NoStop}%
\bibitem [{\citenamefont {Jones}(2015)}]{Jones}%
  \BibitemOpen
  \bibfield  {author} {\bibinfo {author} {\bibfnamefont {R.~O.}\ \bibnamefont
  {Jones}},\ }\href {\doibase 10.1103/RevModPhys.87.897} {\bibfield  {journal}
  {\bibinfo  {journal} {Rev. Mod. Phys.}\ }\textbf {\bibinfo {volume} {87}},\
  \bibinfo {pages} {897} (\bibinfo {year} {2015})}\BibitemShut {NoStop}%
\bibitem [{\citenamefont {Savin}\ \emph {et~al.}(1998)\citenamefont {Savin},
  \citenamefont {Umrigar},\ and\ \citenamefont {Gonze}}]{Gonze}%
  \BibitemOpen
  \bibfield  {author} {\bibinfo {author} {\bibfnamefont {A.}~\bibnamefont
  {Savin}}, \bibinfo {author} {\bibfnamefont {C.~J.}\ \bibnamefont {Umrigar}},
  \ and\ \bibinfo {author} {\bibfnamefont {X.}~\bibnamefont {Gonze}},\
  }\href@noop {} {\bibfield  {journal} {\bibinfo  {journal} {Chemical Physics
  Letters}\ }\textbf {\bibinfo {volume} {288}},\ \bibinfo {pages} {391}
  (\bibinfo {year} {1998})}\BibitemShut {NoStop}%
\bibitem [{\citenamefont {Stowasser}\ and\ \citenamefont
  {Hoffmann}(1999)}]{Stowasser}%
  \BibitemOpen
  \bibfield  {author} {\bibinfo {author} {\bibfnamefont {R.}~\bibnamefont
  {Stowasser}}\ and\ \bibinfo {author} {\bibfnamefont {R.}~\bibnamefont
  {Hoffmann}},\ }\href@noop {} {\bibfield  {journal} {\bibinfo  {journal} {J.
  Amer. Chem. Soc.}\ }\textbf {\bibinfo {volume} {121}},\ \bibinfo {pages}
  {3414} (\bibinfo {year} {1999})}\BibitemShut {NoStop}%
\bibitem [{\citenamefont {Martin}(2004)}]{Martin}%
  \BibitemOpen
  \bibfield  {author} {\bibinfo {author} {\bibfnamefont {R.~M.}\ \bibnamefont
  {Martin}},\ }\href@noop {} {\emph {\bibinfo {title} {Electronic structure:
  Basic theory and practical methods}}}\ (\bibinfo  {publisher} {Cambridge
  university press},\ \bibinfo {year} {2004})\BibitemShut {NoStop}%
\bibitem [{\citenamefont {Perdew}\ \emph {et~al.}(1982)\citenamefont {Perdew},
  \citenamefont {Parr}, \citenamefont {Levy},\ and\ \citenamefont
  {Balduz}}]{Perdew}%
  \BibitemOpen
  \bibfield  {author} {\bibinfo {author} {\bibfnamefont {J.~P.}\ \bibnamefont
  {Perdew}}, \bibinfo {author} {\bibfnamefont {R.~G.}\ \bibnamefont {Parr}},
  \bibinfo {author} {\bibfnamefont {M.}~\bibnamefont {Levy}}, \ and\ \bibinfo
  {author} {\bibfnamefont {J.~L.}\ \bibnamefont {Balduz}},\ }\href {\doibase
  10.1103/PhysRevLett.49.1691} {\bibfield  {journal} {\bibinfo  {journal}
  {Phys. Rev. Lett.}\ }\textbf {\bibinfo {volume} {49}},\ \bibinfo {pages}
  {1691} (\bibinfo {year} {1982})}\BibitemShut {NoStop}%
\bibitem [{\citenamefont {Levy}\ \emph {et~al.}(1984)\citenamefont {Levy},
  \citenamefont {Perdew},\ and\ \citenamefont {Sahni}}]{Levy}%
  \BibitemOpen
  \bibfield  {author} {\bibinfo {author} {\bibfnamefont {M.}~\bibnamefont
  {Levy}}, \bibinfo {author} {\bibfnamefont {J.~P.}\ \bibnamefont {Perdew}}, \
  and\ \bibinfo {author} {\bibfnamefont {V.}~\bibnamefont {Sahni}},\ }\href
  {\doibase 10.1103/PhysRevA.30.2745} {\bibfield  {journal} {\bibinfo
  {journal} {Phys. Rev. A}\ }\textbf {\bibinfo {volume} {30}},\ \bibinfo
  {pages} {2745} (\bibinfo {year} {1984})}\BibitemShut {NoStop}%
\bibitem [{\citenamefont {Janak}(1978)}]{Janak}%
  \BibitemOpen
  \bibfield  {author} {\bibinfo {author} {\bibfnamefont {J.~F.}\ \bibnamefont
  {Janak}},\ }\href {\doibase 10.1103/PhysRevB.18.7165} {\bibfield  {journal}
  {\bibinfo  {journal} {Phys. Rev. B}\ }\textbf {\bibinfo {volume} {18}},\
  \bibinfo {pages} {7165} (\bibinfo {year} {1978})}\BibitemShut {NoStop}%
\end{thebibliography}%

\end{document}